\documentclass[prd,12pt,nofootinbib,preprint]{revtex4}
\usepackage[T1]{fontenc}
\usepackage[utf8x]{inputenc}
\usepackage{amsmath,amssymb}
\usepackage{url}
\usepackage{subfigure}
\usepackage{graphicx}
\usepackage{grffile}
\usepackage[usenames,dvipsnames]{color}
\usepackage{mathtools}
\usepackage{slashed}
\usepackage{array}
\usepackage{multirow}
\usepackage{xcolor}
\usepackage[colorlinks,citecolor=blue]{hyperref}
\usepackage{color}
\usepackage{cancel}
\usepackage{gensymb}
\usepackage{soul}
\usepackage{tikz}
\usepackage{tikz-feynman}

\def\e {\epsilon}

\def\bar {\overline}

\def\be {\begin{equation}}
\def\ee {\end{equation}}
\def\beq {\begin{equation}}
\def\eeq {\end{equation}}
\def\bea {\begin{eqnarray}}
\def\eea {\end{eqnarray}}

\newcommand{\besub}{\begin{subequations}}
\newcommand{\eesub}{\end{subequations}}

\def\beq{\begin{equation}}
\def\eeq{\end{equation}}
\def\barr{\begin{array}}
\def\earr{\end{array}}

\def\e{\epsilon}

\def\q2 {q^2}

\def\bt{\begin{table}}
\def\et{\end{table}}

\begin{document}

\title{Non-resonant Flavoured leptogenesis with aligned two-Higgs doublet model and right-handed neutrinos}

\author{Km Pranjali}
\email{2504110003@mail.jiit.ac.in}
\affiliation{Department of Physics and Material Science and Engineering,
Jaypee Institute of Information Technology,
A-10, Sector-62, Noida 201307, Uttar Pradesh, India}
\author{Indrani Chakraborty}
\email{indrani.chakraborty@mail.jiit.ac.in, indrani300888@gmail.com}
\affiliation{Department of Physics and Material Science and Engineering,
Jaypee Institute of Information Technology, A-10,
Sector-62, Noida 201307, Uttar Pradesh, India}

\vspace{5mm}

\begin{abstract} 

We explore the possibility of non-resonant flavoured leptogenesis in the framework of the aligned two-Higgs doublet model augmented with two right-handed neutrinos. Considering the lightest heavy neutrino mass between 10 TeV-100 TeV, we calculate the baryon asymmetry of the universe by solving fully flavoured coupled Boltzmann equations and find that the observed baryon asymmetry can be
successfully generated for lightest heavy-neutrino masses in this range without relying on resonant
enhancement. We compute the flavour-dependent CP asymmetries and washouts as well. Unlike conventional analyses based on only one Higgs doublet, in this framework, both Yukawa structures associated with the scalar doublets contribute to the CP asymmetry, and their combined effect plays an essential role in realizing the observed lepton asymmetry. We find that the flavoured lepton asymmetry survives if the washout in that direction is weak.  The CP asymmetries in the other flavour directions suffer strong washout. The surviving flavour-dependent lepton asymmetry becomes the main contributor to the final baryon asymmetry. We perform a comprehensive numerical analysis to identify the allowed regions of the multi-dimensional parameter space compatible with neutrino oscillation data, perturbativity of the Yukawa couplings, charged-lepton flavour violation and the observed baryon asymmetry.

\end{abstract} 
\maketitle

\section{Introduction}
\label{intro}

The experimentally observed matter-antimatter asymmetry being one of the most fascinating puzzles of particle physics and cosmology, it can be concluded that there exists a tiny and non-zero excess of baryons over anti-baryons in the universe\cite{DiBari:2021fhs,Davidson:2008bu}. It is measured experimentally by the {\em Planck} collaboration and the aforementioned asymmetry is parametrised in terms of baryon-to-photon ratio \cite{Planck:2018vyg}:
\bea
\eta_B \equiv \frac{n_B - n_{\bar{B}}}{n_\gamma}=(6.12\pm0.04)\times10^{-10}\,,
\eea
where $n_B, n_{\bar{B}}$ and $n_\gamma$ are the number densities of baryons, anti-baryons and photons respectively. The dynamical generation of this asymmetry requires the fulfillment of three Sakharov conditions \cite{Davidson:2008bu,Riotto:1999yt,Sakharov:1967dj}, {\em i.e.} baryon number violation, C or CP violation and out-of-equilibrium conditions. Although all these three conditions are satisfied within the ambit of Standard Model (SM), the generated CP violation is insufficient to generate the observed value of the baryon asymmetry. Thus the baryon asymmetry generated within SM is many orders of magnitude smaller than the observed value. This necessitates an appropriate extension of SM, {\em i.e.} beyond Standard Model (BSM) framework. Furthermore the SM cannot explain the observed non-zero light neutrino masses. Thus BSM scenarios embedding different types of see-saw mechanisms become inevitable for the light neutrino mass generation.

Type-I see-saw mechanism  \cite{Davidson:2008bu,DiBari:2012fz,Sawada:1979dis,Minkowski:1977sc,Mohapatra:1979ia,Babu:1993qv,King:2003jb,Ma:2006km} incorporates additional heavy right-handed neutrinos (RHNs) along with the SM for active neutrino mass generation. The out-of-equilibrium decays of the lightest heavy neutrino  produce lepton asymmetry via CP-violating Yukawa interactions. This lepton asymmetry is further converted into the baryon asymmetry via $(B+L)$-violating electroweak sphaleron processes \cite{Fukugita:1986hr,Kuzmin:1985mm,Buchmuller:2005eh,Buchmuller:2002rq}. This entire procedure of converting lepton asymmetry into baryon asymmetry is termed as {\em leptogenesis}, which is one of the most attractive scenarios for generating the observed baryon asymmetry. Thus the leptogenesis provides a unified framework connecting the origins of active neutrino masses and baryon asymmetry generation.

The complex Yukawa couplings involved in the out-of-equilibrium decay of the lightest heavy neutrino provide one of the ingredients of CP violation. For heavy neutrino masses below $\sim 10^9$ GeV, the charged lepton Yukawa couplings for all three flavours enter the thermal equilibrium. As a result, the quantum coherence between three flavours is lost and this requires a flavour-dependent treatment of the generated asymmetries and washouts. Thus both CP-asymmetries and washouts for each flavour become distinct and the leptogenesis turns out to be flavoured one. 

One of the minimal extensions of SM is two-Higgs doublet models (2HDM) \cite{Branco:1980sz,Branco:2011iw,Gunion:1989we,Bhattacharyya:2015nca,Branco:1999fs}, where the SM is augmented with one extra doublet. These models are phenomenologically rich owing to their extensive usage in studying supersymmetry \cite{Branco:2011iw,Haber:1984rc}, collider physics, several flavour observables, CP violation, etc. A generic 2HDM, by construction can lead to flavour changing neutral currents (FCNCs), which is largely constrained by experimental bounds \cite{Botella:2014ska,Abada:2006ea,Abada:2006fw,Dev:2017trv,Samanta:2019yeg,Blanchet:2006be,Branco:2011iw}. 

In this work, for successful generation of baryon asymmetry, we consider {aligned two Higgs doublet model} (A2HDM) \cite{Celis:2013rcs,Pich:2009sp,Tuzon:2010vt,Karan:2023kyj,Karan:2024kgr,Jung:2010ik}, one of the variants of 2HDMs, along with two right-handed neutrinos (RHNs). Although the leptogenesis in minimal type-I see-saw framework has been largely investigated with different extended scalar sectors \cite{Branco:2011iw,Gunion:2002zf}, this description produces a unified framework for generating the active neutrino mass through type-I see-saw mechanism, charged lepton flavour violation and observed baryon asymmetry of the universe via flavoured leptogenesis. This model nicely eliminates the FCNC by requiring the Yukawa coupling matrices coupled with two doublets to be aligned in flavour space, without applying any discrete symmetries \cite{Branco:2011iw}. To incorporate the alignment, complex alignment parameters need to be introduced, which provides additional sources of CP violation, after remaining consistent with the current experimental bounds. The complex Yukawa couplings corresponding to two different doublets are computed using Casas-Ibarra (CI) parametrisation to ensure the consistency with neutrino oscillation data \cite{Casas:2001sr,Ibarra:2003up,DiBari:2012fz}. We investigate the evolution of the lightest heavy neutrino abundance and the individual lepton flavour asymmetries by solving coupled Boltzmann equations. To be precise, we incorporate the effects of decay, inverse decay, $\Delta L = 1$ scattering and spectator effects as well.  The CP asymmetry surviving the weak washout, contributes dominantly to the finally generated $(B-L)$ symmetry, which is converted into baryon asymmetry through sphaleron processes. We investigate the multi-dimensional parameter space, which is compatible with relevant theoretical and experimental constraints. Through an exhaustive numerical analysis, the main question addressed in this work is whether the additional Yukawa structure of the A2HDM can facilitate successful non-resonant flavoured leptogenesis at the TeV scale.

In a nutshell, the main prospects of our analysis are as follows:
\begin{itemize}
    \item Successful baryogenesis can be achieved with the heavy neutrino mass range $10$ TeV-$100~\mathrm{TeV}$ through non-resonant flavoured leptogenesis.
    
    \item The complex scale factor used for the alignment of two Yukawa couplings provides an additional source of CP-violation. 
    
    \item The interplay between
    flavoured CP asymmetries and hierarchical washouts plays a crucial role in determining the final baryon asymmetry. The CP-asymmetry surviving through a weak washout in a particular flavour direction is retained and contributes majorly in the computation of baryon asymmetry.
    
\item The parameter points yielding the observed baryon asymmetry are
predominantly associated with relatively small heavy-neutrino mass
splittings. Despite this correlation, the mass splittings remain
sufficiently larger than the corresponding decay-width scale,      keeping the successful parameter points in the non-resonant regime.
\end{itemize}    

The paper is structured as follows. In Sec.~\ref{Model}, we present a brief overview of the aligned two-Higgs doublet model with two right-handed neutrinos and discuss the neutrino mass generation mechanism in Sec.~\ref{nu-mix}. The relevant theoretical and experimental constraints imposed on the model are summarized in Sec.~\ref{Constraints}. The formalism of flavoured leptogenesis and the coupled Boltzmann equations are presented in Sec.~\ref{leptogenesis}.  The numerical results and phenomenological implications are discussed in Sec.~\ref{results}. Finally, we summarise and conclude in Sec.~\ref{conclusion}.

\section{Model}
\label{Model}


In this work, we consider the aligned two-Higgs doublet model (A2HDM) with an additional scalar $SU(2)_L$ doublet with hyper-charge $Y=1/2$ (apart from the SM Higgs doublet), augmented with two heavy right-handed neutrinos (RHNs). The additional singlet heavy neutrinos generate light neutrino masses through the type-I see-saw mechanism and simultaneously provide the ingredients required for leptogenesis. In this model, the Yukawa coupling matrices associated with two scalar doublets are aligned in flavour space in order to avoid tree-level flavour-changing neutral current (FCNC). Thus this Yukawa alignment eliminates the tree-level FCNC without imposing any discrete $Z_2$-symmetry.

Two-Higgs doublets can be written as,

\begin{equation}
\Phi_i=
\begin{pmatrix}
\phi_i^+\\
\dfrac{1}{\sqrt2}(v_i+\rho_i+i\eta_i)
\end{pmatrix},
\qquad i=1,2.
\end{equation}

Here, $v_1$ and $v_2$ denote the vacuum expectation values (VEVs) of the neutral components, satisfying: $v=\sqrt{v_1^2+v_2^2}=246~\mathrm{GeV}$. We define $\tan \beta$ as $\tan\beta=\frac{v_2}{v_1}.$

The scalar spectrum of the model comprises of two CP-even neutral Higgs bosons ($h,H$), one CP-odd neutral Higgs boson ($A$), and a pair of charged Higgs bosons ($H^\pm$). The CP-even neutral fields can be rotated to the weak eigenstates through the mixing angle $\alpha$, whereas the CP-odd and charged fields are obtained through rotations involving the angle $\beta$. The details of the scalar sector of the framework can be found in \cite{Branco:2011iw}. We refrain ourselves presenting it again to avoid repetitions. 

\subsection{Yukawa sector}

In this present work, we generate the active neutrino masses via type-I see-saw mechanism by adding two heavy gauge-singlet RHNs $N_i$ ($i=1,2$) with the aligned two Higgs doublet model. Thus we only write down the relevant part of the Yukawa Lagrangian involving the RHNs as follows:

\begin{equation}
\mathcal{L}_{N}
=
-\sum_{\alpha=e,\mu,\tau}\sum_{i=1}^{2}
\bar{L}_{L\alpha}
\left[
Y^1_{\alpha i}\,\widetilde{\Phi}_1
+
Y^2_{\alpha i}\,\widetilde{\Phi}_2
\right]
N_{i}
-\frac{1}{2}
\sum_{i,j=1}^{2}
\bar{N_{i}^{\,c}}
(M_R)_{ij}
N_{j}
+\mathrm{h.c.},
\label{eq:NuLag}
\end{equation}
Here, $L_{L\alpha}$ ($\alpha=e,\mu,\tau$) denotes the left-handed lepton doublet, $\widetilde{\Phi}_a=i\sigma_2\Phi_a^\ast$ ($a=1,2$), $Y^1$ and $Y^2$ are the $3\times2$ complex  neutrino Yukawa coupling matrices, and $M_R$ is the $2\times2$ Majorana mass matrix.

Since the RHNs are gauge singlets under the Standard Model gauge group, by performing a unitary transformation in the RHN flavour space, one can diagonalize the complex symmetric Majorana mass matrix.  Therefore, for our analysis, we use the basis, where $M_R$ is diagonal, {\em{i.e.}},
\begin{equation}
M_R=
\mathrm{diag}(M_1,M_2),
\end{equation}
where $M_1$ and $M_2$ are the masses of the two heavy neutrinos. This choice does not alter the physical predictions, since the corresponding phase can be absorbed into a redefinition of the RHNs and Yukawa couplings. In this analysis, from the perspective of the leptogenesis, we have considered $M_2 > M_1.$

As stated previously, the removal of tree-level FCNC is guaranteed by the alignment of the Yukawa coupling matrices in the quark and charged lepton sector, {\em i.e.} we assume that the same mechanism also holds in the neutrino sector as well \cite{DelleRose:2019ukt,Tuzon:2010vt}. Thus the matrices $Y^1$ and $Y^2$ are also proportional to each other in flavour space:
\begin{equation}
Y^2
=
\xi_\nu\,Y^1,
\label{eq:alignment}
\end{equation}
where $\xi_\nu$ is the complex neutrino alignment parameter, which can be decomposed into real and imaginary parts as: $\xi_\nu = {\xi_\nu}_R + I {\xi_\nu}_I $.Consequently, Eq.~(\ref{eq:NuLag}) reduces to  :

\begin{equation}
\mathcal{L}_{N}
=
-\sum_{\alpha=e,\mu,\tau}\sum_{i=1}^{2}
\bar{L}_{L\alpha}
Y^1_{\alpha i}
\left(
\widetilde{\Phi}_1
+
\xi_\nu\widetilde{\Phi}_2
\right)
N_{i}
-\frac{1}{2}
\sum_{i=1}^{2}
M_R\,
\bar{N_{i}^{\,c}}
N_{i}
+\mathrm{h.c.}.
\label{eq:NuLagAligned}
\end{equation}

\section{Neutrino masses and mixing: fitting of neutrino data}
\label{nu-mix}

After electroweak symmetry breaking (EWSB), following Eq.(\ref{eq:NuLagAligned}), the Dirac neutrino mass matrix $M_D$ becomes,
\bea
M_D &=&
Y^1
(\cos\beta+\xi_\nu\sin\beta) \frac{v}{\sqrt2}, \nonumber \\
&=& Y_\nu^{\rm eff}  \frac{v}{\sqrt2}\,.
\label{eff}
\eea

where, $Y_\nu^{\rm eff}
=
Y^1
(\cos\beta+\xi_\nu\sin\beta)$.

We have taken each and every entry of the complex $3 \times 2$ Yukawa coupling matrix $Y^1$ to be complex, which might contribute to the successful source of CP-violation. The active neutrino mass is generated through the type-I see-saw mechanism  owing to the presence of heavy RHNs. Corresponding active neutrino mass matrix can be written as:

\begin{equation}
M_\nu
=
-M_D
M_R^{-1}
M_D^T.
\end{equation}

To generate the light neutrino masses, we diagonalise $M_\nu$ by the following transformation: $U^T_{\rm PMNS}~ M_\nu ~U_{\rm PMNS}  = {\rm diag} ~(m_1, m_2, m_3) = M_\nu^{\rm diag}$. $m_1, m_2, m_3$ are three active neutrino masses and $U_{\rm PMNS}$ is  the Pontecorvo-Maki-Nakagawa-Sakata matrix \cite{Maki:1962mu,Giganti:2017fhf,ParticleDataGroup:2024cfk,Esteban:2024eli}. $U_{\rm PMNS}$ matrix can be written as: 
\begin{equation}
U_{\rm PMNS}
=
\begin{pmatrix}
c_{12}c_{13}
&
s_{12}c_{13}
&
s_{13}e^{-i\delta}
\\[2mm]
-s_{12}c_{23}
-c_{12}s_{23}s_{13}e^{i\delta}
&
c_{12}c_{23}
-s_{12}s_{23}s_{13}e^{i\delta}
&
s_{23}c_{13}
\\[2mm]
s_{12}s_{23}
-c_{12}c_{23}s_{13}e^{i\delta}
&
-c_{12}s_{23}
-s_{12}c_{23}s_{13}e^{i\delta}
&
c_{23}c_{13}
\end{pmatrix},
\label{eq:PMNS}
\end{equation}
where $s_{ij} (c_{ij})$ is $\sin \theta_{ij} (\cos \theta_{ij})$ and $\delta$ is the CP-violating phase. The light neutrino spectrum admits two possible orderings between the active neutrino masses according to their hierarchy: (i) $m_1 < m_2 < m_3$ for normal ordering (NO) and (ii) $m_3 < m_1 < m_2$ for inverted ordering (IO). 

\subsection{Casas-Ibarra parametrization}

We employ the Casas Ibarra (CI) parametrisation \cite{Casas:2001sr,Ibarra:2003up,Xing:2009vb} for writing the Dirac mass matrix $M_D$,
\bea
M_D = i\, U_\mathrm{PMNS}\,
\sqrt{M_\nu^{\text{diag}}}\,
R\,
\sqrt{M_R},
\label{CI-param}
\eea
where $R$ is a $3 \times 2$ complex orthogonal matrix. We can consider one of the active neutrino masses to be zero for simplicity, in that case, $R$ can be written using a single complex angle $z = x + I y$:
\bea
R = 
\begin{pmatrix}
0 & 0 \\
\cos z & -\sin z \\
\sin z & \cos z
\end{pmatrix},
\quad \text{(NO)}, \nonumber 
\eea
and
\bea
R = 
\begin{pmatrix}
\cos z & -\sin z \\
\sin z & \cos z \\
0 & 0
\end{pmatrix},
\quad \text{(IO)}.
\eea

Eq.(\ref{CI-param}) allows us to reconstruct the effective Yukawa matrix using the experimentally measured neutrino oscillation parameters together with the free parameters of the heavy-neutrino sector. We have fixed the $\theta_{12}, \theta_{13}, \theta_{23}$ and $\delta$ at their respective central values obtained from neutrino oscillation data \cite{Esteban:2024eli,ParticleDataGroup:2024cfk}. For simplicity we choose the lightest active neutrino mass to be zero. Therefore the masses of the other active neutrinos are also fixed from the neutrino oscillation data. The effective Yukawa matrix is reconstructed in terms of $U_{\rm PMNS}, M_\nu^{\text{diag}}, R$ and $M_R$ using CI-parametrisation.

\section{Constraints}
\label{Constraints}

The multi-dimensional parameter space of the aligned two Higgs doublet model is restricted by several theoretical and experimental constraints \cite{Karan:2023kyj,Karan:2024kgr,Chakrabarty:2014aya,Branco:2011iw,Davidson:2002qv,Hambye:2003rt,Eberhardt:2020dat}. Before computing the observed baryon asymmetry of the universe, all parameter points need to satisfy these constraints. Throughout our analysis, the scalar-sector parameters are taken in a phenomenologically viable region of the aligned two-Higgs doublet model \cite{Karan:2023kyj,Karan:2024kgr,Chakrabarty:2014aya,Branco:2011iw,Davidson:2002qv,Hambye:2003rt,Eberhardt:2020dat}. In the numerical analysis, we particularly impose the perturbativity of the Yukawa couplings and the bound coming from the lepton flavour violating processes, particularly from $\mu \rightarrow e \gamma$. We also check the consistency with current neutrino oscillation data through previously mentioned CI-parametrisation. Finally we employ the surviving parameter points for baryon asymmetry calculations.
In the following subsections, we shall present the list of constraints applied to the parameter space, to make it a viable one.

\subsection{Perturbativity}

Perturbativity is enforced on all the elements of the Yukawa coupling matrices $Y^1$ and $Y^2$ present in Eq.(\ref{eq:NuLag}) throughout the parameter space. 
 In the present analysis, we require each element of the Yukawa coupling matrices to be sufficiently small and satisfy simultaneously,
\begin{equation}
\left|Y^1_{\alpha i}\right|<\sqrt{4\pi},
\qquad
\left|Y^2_{\alpha i}\right|<\sqrt{4\pi},
\end{equation}
where $\alpha=e,\mu,\tau$ and $i=1,2$. 

\subsection{Neutrino oscillation data and CI-parametrisation}

After EWSB, two individual Yukawa couplings $Y^1$ and $Y^2$ combine with each other to generate the Dirac mass involving the effective Yukawa coupling mentioned in Eq.(\ref{eff}). Further, Dirac mass matrix $M_D$ or $Y_\nu^{\rm eff}$ is expressed in terms of $U_{\rm PMNS}, M_\nu^{\rm diag}$, orthogonal matrix R and $M_R$ through Casas- Ibarra parametrisation. It is not sufficient for the individual Yukawa couplings to be perturbative, but they also have to comply with the observed neutrino oscillation data. 

As mentioned earlier, one of the three active neutrinos is massless. For the analysis, we have checked the results using both NO and IO hierarchy and found that qualitative results remain the same for these two hierarchies. Thus, we present our analysis considering the NO hierarchy in this paper. Throughout the numerical analysis, the neutrino oscillation parameters in $U_{\rm PMNS}$ matrix are fixed at their current central values obtained from the latest global analysis of neutrino oscillation data~\cite{Esteban:2024eli}. For simplicity, we have set the Majorana phase to be zero. The free parameters in the CI-parametrisation are real and imaginary parts ($x$ and $y$) of the complex angle $z$ and the masses of heavy right-handed neutrinos $M_1$ and $M_2$. By varying these free parameters, we shall obtain different values of $Y_\nu^{\rm eff}$. This in turn will produce separate sets of $Y^1$ and $Y^2$ for fixed or different values of $\tan \beta$ and $\xi_\nu$.

\subsection{Charged Lepton Flavour Violation}

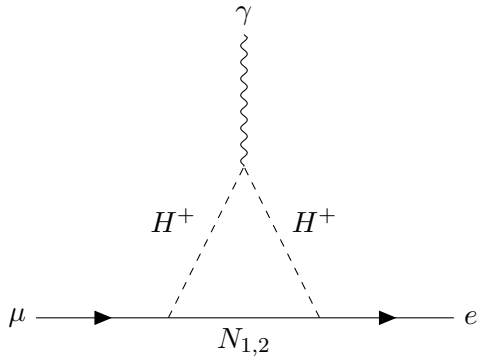
\begin{figure}[h]
\centering

\begin{tikzpicture}
\begin{feynman}

\vertex (a) at (0,0) {\(\mu\)};
\vertex (c) at (2,0);
\vertex (d) at (4,0);
\vertex (e) at (6,0) {\(e\)};

\vertex (f) at (3,2);
\vertex (b) at (3,4) {\(\gamma\)};

\diagram*{
(a) -- [fermion] (c)
    -- [plain, edge label'=\(N_{1,2}\)] (d)
    -- [fermion] (e),

(c) -- [scalar, edge label=\(H^+\)] (f),
(d) -- [scalar, edge label'=\(H^+\)] (f),
(f) -- [photon] (b),
};

\end{feynman}
\end{tikzpicture}

\caption{Feynman diagram of the process $\mu \rightarrow \e \gamma$, mediated by a charged scalar.}
\label{fig:muegamma}

\end{figure}

The alignment of two Yukawa couplings in aligned two Higgs doublet model by construction leads to the elimination of FCNCs. Thus this model alone does not allow for charged lepton flavour violation. But the addition of two heavy right-handed neutrinos gives rise to lepton flavour-violating neutrino Yukawa interactions in radiative decays such as $\mu \rightarrow e\gamma$ at the one-loop level. The one-loop diagram drawn in Fig.\ref{fig:muegamma}, receives additional contributions from the charged Higgs and heavy neutrino interactions, thereby modifying the LFV amplitudes.

Among the various LFV observables, the strongest bound comes from  the decay $\mu \rightarrow e \gamma$. All the parameter points satisfying the above constraints, are required to obey the current experimental upper bound on the branching ratio \cite{MEGII2025},
\begin{equation}
\mathrm{Br}(\mu \rightarrow e\gamma) < 4.2 \times 10^{-13}.
\end{equation}
The analytical expressions for the branching ratio and the relevant Feynman diagrams are presented in Appendix~\ref{A}, following Ref.~\cite{Lindner:2016bgg,Minkowski:1977sc,Paul:2026snc,Cai:2017jrq}. 

\subsection{Baryon Asymmetry of the Universe}

The primary objective of this work is to produce the observed baryon asymmetry through leptogenesis. In this mechanism, lepton asymmetry is generated via CP-violating and out-of-equilibrium decays of the lightest right-handed neutrino $N_1$, which is further converted into a baryon asymmetry via the electroweak sphaleron processes. Therefore one must compute the baryon asymmetry and check whether it is compatible with the current observed data.

The baryon-to-photon ratio measured by the \textit{Planck} collaboration is given by~\cite{Planck:2018vyg}
\begin{equation}
\eta_B = (6.12 \pm 0.04)\times10^{-10}.
\end{equation}

The corresponding baryon asymmetry in terms of the baryon number density normalized to the entropy density is
\begin{equation}
Y_B^{\rm obs} = (8.69 \pm 0.06)\times10^{-11}.
\label{baryon-asy}
\end{equation}

To obtain the final baryon asymmetry for each parameter points, coupled Boltzmann equations need to be solved. The parameter points satisfying the previous theoretical and experimental data, should reproduce the observed value of baryon asymmetry to become phenomenologically viable. For our analysis, we filter those parameter points which lie within $3 \sigma$-error bar of the aforementioned data in Eq.(\ref{baryon-asy}).

Having established relevant theoretical and experimental constraints, we now proceed towards the formalism of flavoured leptogenesis to compute the baryon asymmetry of the universe.

\section{Leptogenesis Formalism}
\label{leptogenesis}
As mentioned previously, in this framework, we consider that the lepton asymmetry is generated dominantly from the out-of-equilibrium decay of the lightest heavy neutrino $N_1$ into SM lepton and Higgs doublet. The lepton asymmetry generated from the decay of the heavier neutrino $N_2$ is erased through strong washouts. The complex Yukawa coupling driving the decay, becomes one of the ingredients of CP-violation. At very high temperature, the heavy RHNs are produced in the thermal bath through inverse decay. With the drop of temperature below the mass of the lightest RHN $N_1$, decay of $N_1$ contributes dominantly in lepton-asymmetry generation. This lepton asymmetry is subsequently converted into baryon asymmetry via $(B+L)$-violating, but $(B-L)$ conserving electroweak {\em sphaleron} processes.

Since the heavy neutrinos masses considered for the lepton asymmetry generation in this analysis lie well below $10^9~\mathrm{GeV}$, the flavour effects need to be incorporated. Therefore we proceed towards the calculation of the flavour dependent CP-asymmetry and derivation of the flavoured Boltzmann equations controlling the evolution of the heavy neutrino abundance and the individual lepton flavour asymmetries, after a brief discussion of flavour effects.

\subsection{Flavour Effects}

The generation of the baryon asymmetry through leptogenesis is highly sensitive to the thermal history of the early Universe. When the purely charged lepton Yukawa interactions become faster than the heavy neutrino interactions, during the out-of-equilibrium decay of the lightest heavy neutrino $N_1$, the charged lepton Yukawa interactions for one or more flavour enter into the equilibrium\cite{Barbieri:1999ma,Vives:2005ra,Branco:2011iw}. As a result, the quantum coherence between different lepton flavours in the plasma is destroyed and they can be distinguished as distinct flavours. Consequently, the leptogenesis dynamics can be classified into distinct flavour regimes.

The charged lepton interactions are out-of-equilibrium at very high temperatures ($T\gtrsim10^{12}\,\mathrm{GeV}$), resulting in the coherence of the lepton flavours. The lepton asymmetry generated in this {\em unflavoured regime} is independent of the individual flavours. The intermediate temperature range,
$10^{9}\,\mathrm{GeV}\lesssim T \lesssim10^{12}\,\mathrm{GeV}$, is termed as {\em two-flavour regime} since the $\tau$-flavour attains the equilibrium, whereas the electron and muon flavours remain coherent. Finally, at temperatures below $10^{9}\,\mathrm{GeV}$ all three flavours can be distinguished as stated previously, initiating the {\em fully flavoured regime}.

In this work, we consider that the heavy neutrino masses lie in the range $\mathcal{O}(10~\mathrm{TeV})$. Therefore the dynamics of leptogenesis will take place in fully flavoured regime. Instead of considering the total flavour independent CP asymmetry, now we have to calculate CP asymmetry separately for each lepton flavour. Corresponding washouts will also be flavour dependent now. This flavour-dependent treatment plays a crucial role in our analysis, as we shall uncover later.

\subsection{Flavoured CP Asymmetry}

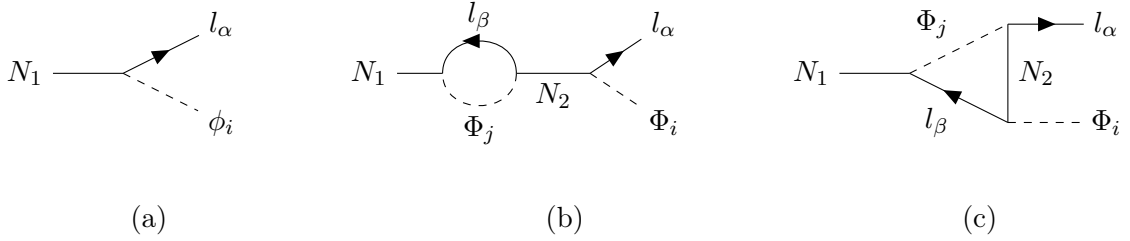
\begin{figure}[h]
\centering

\begin{tikzpicture}[scale=0.65]

\begin{feynman}

\vertex (a1) at (0,0) {\(N_{1}\)};
\vertex (b1) at (2,0);
\vertex (c1) at (4,1) {\(l_\alpha\)};
\vertex (d1) at (4,-1) {\(\phi_i\)};

\diagram*{
(a1)--[plain](b1),
(b1)--[fermion](c1),
(b1)--[scalar](d1),
};

\end{feynman}

\begin{feynman}

\vertex (a2) at (7,0) {\(N_{1}\)};
\vertex (p2) at (8.5,0);
\vertex (n2) at (10,0);
\vertex (b2) at (11.5,0);

\vertex (c2) at (13,-1) {\(\Phi_i\)};
\vertex (d2) at (13,1) {\(l_\alpha\)};

\diagram*{
(a2)--[plain](p2),
(p2)--[anti fermion, half left, edge label=\(l_\beta\)](n2),
(n2)--[scalar, half left, edge label=\(\Phi_j\)](p2),
(n2)--[plain, edge label'=\(N_{2}\)](b2),
(b2)--[scalar](c2),
(b2)--[fermion](d2),
};

\end{feynman}

\begin{feynman}

\vertex (a3) at (16,0) {\(N_{1}\)};
\vertex (b3) at (18,0);
\vertex (c3) at (20,1);
\vertex (d3) at (20,-1);

\vertex (e3) at (22,1) {\(l_\alpha\)};
\vertex (f3) at (22,-1) {\(\Phi_i\)};

\diagram*{
(a3)--[plain](b3),
(b3)--[anti fermion, edge label'=\(l_\beta\)](d3),
(b3)--[scalar, edge label=\(\Phi_j\)](c3),
(d3)--[scalar](f3),
(c3)--[fermion](e3),
(d3)--[plain, edge label'=\(N_{2}\)](c3),
};

\end{feynman}

\end{tikzpicture}

\vspace{2mm}

(a)\hspace{5cm}(b)\hspace{5cm}(c)

\caption{The CP asymmetry in Type-I see-saw leptogenesis results from the interference between the tree-level diagram and the one-loop self-energy and vertex diagrams. Here $\Phi_i$ and $\Phi_j$ denote $\Phi_1$ and (or) $\Phi_2$.}

\label{fig:cp}

\end{figure}

The lepton asymmetry generated during leptogenesis originates from the CP-violating decays of the lightest heavy neutrino $N_1$. The interference between the tree-level decay amplitude (Fig.\ref{fig:cp}(a)) and the corresponding one-loop vertex (Fig.\ref{fig:cp}(c)) and self-energy corrections (Fig.\ref{fig:cp}(b)) brings about a non-vanishing CP asymmetry \cite{Fong:2012buy}. In the fully flavoured regime as mentioned previously, the CP asymmetry has to be calculated individually for each and every lepton flavour.


The flavour-dependent CP asymmetry generated in the decay of the  $N_1$ into the charged lepton flavour $\alpha$ ($\alpha = e , \mu, \tau$) and scalar $\phi$ is defined as:
\begin{equation}
    \epsilon_{1\alpha} \equiv 
    \frac{\Gamma(N_1 \to \Phi \ell_\alpha) - \Gamma(N_1 \to \bar\Phi ~\bar\ell_\alpha)}
    {\Gamma(N_1 \to \Phi \ell) + \Gamma(N_1 \to \bar\Phi~ \bar\ell)}.
    \end{equation}
    Unlike the conventional type-I see-saw scenario, where the heavy Majorana neutrinos couple to a single Higgs doublet, in the present framework, $N_1$ couples with two doublets $\Phi_1$ and $\Phi_2$. Thus we need to consider both tree level decays $N_1 \rightarrow \Phi_1 
 \ell_\alpha$ and $N_1 \rightarrow \Phi_2 
 \ell_\alpha$ (Fig.\ref{fig:cp}(a)) along with their self energy and vertex correction counterparts (Fig.\ref{fig:cp}(b)) and (Fig.\ref{fig:cp}(c)) respectively for computing the flavour-dependent CP asymmetry. Corresponding expression of flavoured CP asymmetry $\epsilon_{i \alpha}$ (with $i=1$) can therefore be written as: 
  
\begin{equation}
\e_{1\alpha} = \e^{vertex}_{1\alpha} +\e^{self-energy}_{1\alpha}
\label{epsilon}
\end{equation}
where,\bea \e^{vertex}_{1\alpha}&=& \Bigg[\frac{1}{8\pi}\frac{1}{(Y^{1\dagger }Y^1)_{11}}\left(
\mathrm{Im}\!\left[(Y^{1\dagger} Y^1)_{21} \, Y^1_{\alpha1} Y^{1*}_{\alpha2}\right]+
\mathrm{Im}\!\left[(Y^{1\dagger} Y^2)_{21} \, Y^1_{\alpha1} Y^{2*}_{\alpha2}\right]\right) \nonumber \\
&&
+ \frac{1}{8\pi}\frac{1}{(Y^{2\dagger }Y^2)_{11}}\left(
\mathrm{Im}\!\left[(Y^{2\dagger} Y^1)_{21} \, Y^2_{\alpha1} Y^{1*}_{\alpha2}\right]+
\mathrm{Im}\!\left[(Y^{2\dagger} Y^2)_{21} \, Y^2_{\alpha1} Y^{2*}_{\alpha2}\right]\right)\Bigg]~ g\!\left(\frac{M_2^2}{M_1^2}\right)\nonumber,
\eea

\bea
\e^{self-energy}_{1\alpha}&=& \Bigg[\frac{1}{8\pi}\frac{1}{(Y^{1\dagger }Y^1)_{11}}\left(
\mathrm{Im}\!\left[(Y^{1\dagger} Y^1)_{12} \, Y^1_{\alpha1} Y^{1*}_{\alpha2}\right]+
\mathrm{Im}\!\left[(Y^{2\dagger} Y^2)_{12} \, Y^1_{\alpha1} Y^{1*}_{\alpha2}\right]\right)\nonumber \\
&&
+\frac{1}{8\pi}\frac{1}{(Y^{2\dagger }Y^2)_{11}}\left(
\mathrm{Im}\!\left[(Y^{1\dagger} Y^1)_{12} \, Y^2_{\alpha1} Y^{2*}_{\alpha2}\right]+
\mathrm{Im}\!\left[(Y^{2\dagger} Y^2)_{12} \, Y^2_{\alpha1} Y^{2*}_{\alpha2}\right]\right)\Bigg] \frac{M_1^2}{M_1^2 - M_2^2}\nonumber
\eea

and the loop function $g(x)$ is given by $g(x) = \sqrt{x}
\left[
\frac{1}{1 - x}
+ 1 - (1 + x)\ln\!\left(\frac{1 + x}{x}\right)
\right]$, with
$x=M_2^2/M_1^2$.

From Eq.(\ref{epsilon}), it is clear that the presence of the two Yukawa structures introduces interference
terms involving $Y^1$ and $Y^2$, in addition to the conventional
terms involving a single Yukawa structure. Therefore the flavoured
CP asymmetries are modified accordingly.
This flavoured CP asymmetries will enter into the Boltzmann equations and contribute as source terms. Besides there will be flavour-dependent washout terms too. The evolution and interplay of these two types of terms will determine the final baryon asymmetry of the universe, which will be discussed in the next subsection.

\subsection{Boltzmann equations}
\label{BE}

In this subsection, we solve the Boltzmann equations for computing the baryon asymmetry of the universe in the context of flavoured leptogenesis. The source terms for flavoured leptogenesis in the coupled Boltzmann equations, contain the flavour-dependent CP asymmetries generated in the out-of-equilibrium decay of the lightest heavy neutrino $N_1$. There is a competition between the source terms and the washout terms arising due to inverse decays and $\Delta L=1$ scattering interactions. The interplay between these two terms finally leads to the actual determination of baryon asymmetry. Since we consider the flavoured leptogenesis at a temperature $< 10^9$ GeV, we need to evolve $\Delta_\alpha = \frac{B}{3}-L_\alpha$ with $\alpha = e, \mu, \tau$. For accuracy, we take into account the spectator effects as well. As stated earlier, we only consider the lepton asymmetry generated from the decay of $N_1$.

It is convenient to express the Boltzmann equations in terms of the dimensionless variable $z=\frac{M_1}{T}$, where $T$ denotes the temperature of the thermal plasma. The particle abundances are also normalized to the entropy density, {\em i.e.}
$Y_X=\frac{n_X}{s}$, $n_X$ and $s$ being the number density of the particle species $X$ and entropy density $s$ \footnote{$s$ is defined as $s=\frac{2\pi^2}{45}g_*T^3$, where $g_*$ represents the effective number of relativistic degrees of freedom.}.

The equilibrium abundance of the lightest heavy neutrino and relativistic lepton doublet are given by,
\begin{equation}
Y_{N_1}^{\text{eq}}(z) = \frac{45}{2\pi^4 g_*} z^2 K_2(z)
\end{equation}
\begin{equation}
Y_\ell^{\text{eq}} = \frac{135\, g\, \zeta(3)}{8 \pi^4 \, g_*}
\end{equation}
$g$ being the number of degrees of freedom of the corresponding particles and $K_2$ being second modified Bessel function of second kind \cite{Datta:2023pav,Nardi:2006fx,Buccella:2018jij}.

The evolutions of $N_1$ abundance and individual flavour asymmetries are governed by the following coupled Boltzmann equations:

\begin{subequations}
\bea
\frac{dY_{N_1}}{dz}
&=&
-\frac{1}{sH(M_1)z}
\left(\frac{Y_{N_1}}{Y^{\mathrm{eq}}_{N_1}} - 1\right)
\left(\gamma_D + 2\gamma_{Ss} + 4\gamma_{St}\right) \,,  
\label{BE1}\\
\frac{dY_{\Delta_\alpha}}{dz}
&=&
\frac{1}{sH(M_1)z}
\Bigg[
\left(\frac{Y_{N_1}}{Y_{N_1}^{\mathrm{eq}}} - 1\right)\epsilon_{1 \alpha}\, \gamma_D
- K_{1\alpha} \sum_{\beta=e,\mu,\tau}
\left(
\frac{1}{2}(C^\ell_{\alpha\beta} + C^H_\beta)\gamma_D
\right)\frac{Y_{\Delta_\beta}}{Y^{\mathrm{eq}}_l} \nonumber \\
&&
- K_{1\alpha} \sum_{\beta=e,\mu,\tau}
\left(
 \left(\frac{Y_{N_1}}{Y_{N_1}^{\mathrm{eq}}} - 1\right)\left(C^\ell_{\alpha\beta}\gamma_{Ss}+\frac{C^H_\beta}{2}\gamma_{St}\right)
\right)\frac{Y_{\Delta_\beta}}{Y^{\mathrm{eq}}_l} \nonumber \\
&&
- K_{1\alpha} \sum_{\beta=e,\mu,\tau}
\left( (2C^\ell_{\alpha\beta} + C^H_\beta)\left(\gamma_{St} + \frac{1}{2}\gamma_{Ss}\right)
\right)
\frac{Y_{\Delta_\beta}}{Y^{\mathrm{eq}}_l}
\Bigg].
\label{BE2}
\eea
\end{subequations}
In the above equations, $H(M_1) = 1.66~ g^{* 1/2} {\frac{T^2}{M_{\rm Pl}}|_{T=M_1}}$ is the Hubble's constant at $T=M_1$, $M_{\rm Pl}$ being the Planck scale.
In Eq.(\ref{BE2}), $Y_{\Delta_\alpha}$ ($\alpha=e,\mu,\tau$) denotes the $B/3-L_\alpha$ asymmetry normalized to the entropy density. The first term at the right hand side contributes to the production of lepton asymmetry through the out-of-equilibrium decay of $N_1$, {\em i.e.} the decay rate $\gamma_D$. The remaining terms account for the $\Delta L=1$ Higgs-mediated scattering interaction rates through the $s$- and $t$-channel ($\gamma_S$ and $\gamma_t$) diagrams. The inverse decay rate for the washout mechanism is same as the decay rate $\gamma_D$.  Since all the charged-lepton Yukawa interactions are in thermal equilibrium in the relevant temperature regime, washouts need to be considered for individual flavours. Flavour dependent washouts are denoted by the flavour projector $K_{1\alpha}$ which is defined as: $K_{1\alpha} = \frac{(Y^{1*})_{\alpha 1} (Y^{1})_{\alpha 1} + (Y^{2*})_{\alpha 1} (Y^{2})_{\alpha 1}}{(Y^{1 \dag}Y^1)_{11} + (Y^{2 \dag}Y^2)_{11}}$\cite{Datta:2023pav}. The matrices $C^\ell$ and $C^H$ incorporate the spectator effects, which connect the asymmetry in the Higgs and lepton sector with the asymmetries in $\Delta_\alpha$. Detailed formulae of $\gamma_D, \gamma_S, \gamma_t$, $C^\ell$ and $C^H$ matrices are relegated to Appendix~\ref{B}, following Ref.~\cite{Pramanick:2024gvu,Datta:2023pav,Nardi:2006fx,Giudice:2003jh,Plumacher:1996kc}

Now we need to numerically solve a system of four coupled differential equations subject to the initial conditions, $Y_{N_1}(0) = Y_{N_1}^{eq}$ and $Y_{\Delta_\alpha}(0) = 0$ at $ T >> M_1$. As a result, we obtain the lightest heavy neutrino abundance and the three independent lepton flavour asymmetries as solutions, finally which leads to the total $B-L$ symmetry, {\i.e.} $Y_{B-L} = \sum_\alpha Y_{\Delta_\alpha}$. This $B-L$ symmetry is finally converted to baryon asymmetry through electroweak sphaleron processes according to \cite{Khlebnikov:1988sr}:
\begin{equation}
Y_B
=
\frac{8N_f+4N_H}{22N_f+13N_H}
\,Y_{B-L},
\label{BA}
\end{equation}

where $N_f$ and $N_H$ are the number of fermion generations and Higgs doublets, respectively. In the present work, we substitute, $N_f=3$ and $N_H=2$.

\section{Results and discussions}
\label{results}

\subsection{Numerical analysis}

Next we identify the regions of a multi-dimensional parameter space of the aligned two-Higgs doublet model with two right-handed neutrinos, which lead to the successful generation of observed baryon asymmetry. Thus an exhaustive numerical scan of the aforementioned parameter space is performed with $10~\mathrm{TeV}$
intervals of the mass of $N_1$, over the following relevant parameters within the appropriate ranges:
\bea
 && 10 ~ {\rm TeV}\leq M_1, M_2 \leq 100~ {\rm TeV},~~~ -\pi \leq x,y \leq +\pi, \nonumber \\
 && -1.0 \leq \xi_{\nu R},\xi_{\nu I} \leq 1.0,~~~ 10.0 \leq \tan \beta \leq 30.0,
 \eea
 with $M_2 > M_1$ to ensure successful leptogenesis through the out-of-equilibrium decay of the lightest heavy neutrino $N_1$. This interval-based scanning
strategy enables us to perform a fine scan over the parameters by incorporating higher precision. 
 
First, we filter the parameter points which satisfy the theoretical and experimental constraints mentioned in Sec.~\ref{Constraints}. The points violating any one of these constraints are discarded for the next part of the analysis. The points satisfying all the constraints are used further for the leptogenesis analysis. With these surviving parameter points, the coupled flavoured Boltzmann equations described in sub-section~\ref{BE} are simultaneously solved, incorporating the effects of decays, inverse decays and $\Delta L=1$ scatterings. Since the mass of the heavy right-handed neutrinos lie between 10 TeV-100 TeV, while solving the equations, we also take spectator effects \cite{Nardi:2006fx} into account. The final baryon asymmetry is computed from the resulting flavour asymmetries following Eq.(\ref{BA}). The parameter points within the $3 \sigma$ error bar around the central value of the observed baryon asymmetry \cite{Planck:2018vyg} are considered to be the allowed points.


\subsection{Flavour dependent evolution of lepton asymmetry}

\begin{figure}[htpb!]{\centering
\includegraphics[height = 10 cm, width = 12 cm]{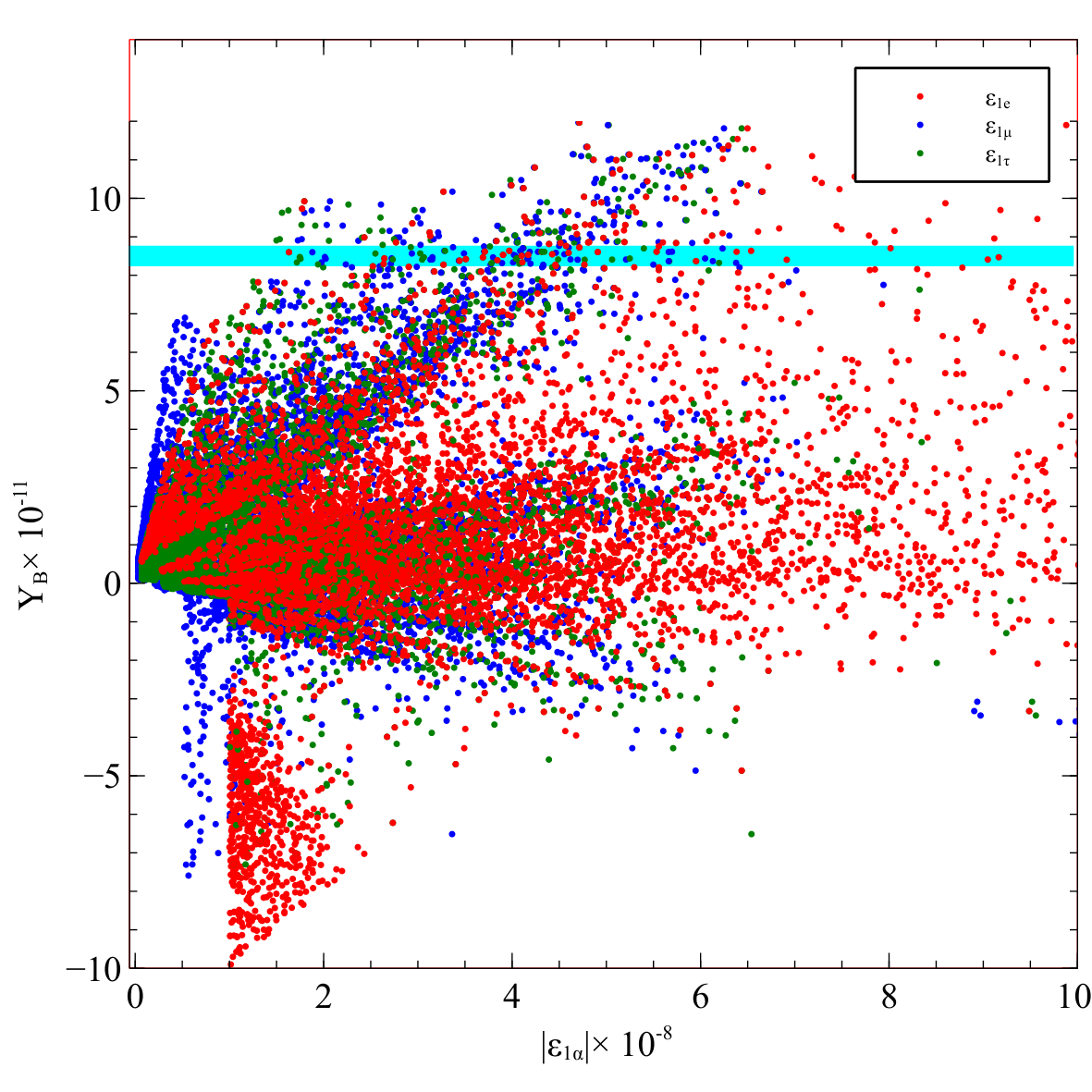}}
\caption{Variations of $Y_B$ with $|\epsilon_\alpha|$, with $\alpha = e, \mu, \tau$. Colour coding is explained in legends.}
\label{YB-vs-mod-ealpha}
\end{figure}

Since the masses of the heavy right-handed neutrinos considered in our analysis are less than $10^9~{\rm GeV}$, three charged lepton flavours $e,\mu$ and $\tau$ are well separated in the thermal bath. In Fig.\ref{YB-vs-mod-ealpha} \footnote{While plotting all the figures, we combine the results from all the scan intervals for getting an overview of the full parameter space which is compatible with the successful leptogenesis.}, we have depicted the variation of $Y_B$ vs. $|\epsilon_{1 \alpha}|$. From the aforementioned evolutions, it is very clear that the CP asymmetries are generated in all flavour directions, but the contributions of the different flavours to the final lepton asymmetry are not equal. The reason for the substantial scatter is the interplay between the generation of flavoured lepton asymmetry and flavoured washouts to obtain the final baryon asymmetry. Thus in principle, sizeable values of $Y_B$ might be generated from moderate values of $\epsilon_{1 \alpha}$. Since the three different flavour sectors occupy different parts of the parameter space, the flavoured Boltzmann equations have distinct contributions from three flavoured CP-asymmetries.

Here, we have plotted the points with negative $Y_B$ also, which comes as an artifact of the flavour-dynamics, {\em i.e.} the mutual cancellations of $Y_{\Delta_\alpha}$'s, which will be discussed in the next subsections. These points having negative $Y_B$, signify the excess of anti-baryons with respect to the baryons, which is forbidden from experimental results. The parameter points for which positive $Y_B$ is produced with magnitude less that the observed value, are termed as "under-abundant" points. On the other hand, if the produced baryon asymmetry is positive, but greater than the observed value, the points correspond to "over-abundant" regime and are discarded by the experimental data.

\subsection{Hierarchical and flavour-dependent washouts}

A large CP-asymmetry in one flavour direction does not indicate a large share in the final baryon asymmetry. The entire parameter space is designed in such a way that for a particular benchmark point, the washout in any one of the three flavour directions is weak and washouts in the other two flavour directions are strong enough to erase the generated CP-asymmetry in that direction. During the final computation of the baryon asymmetry, the surviving flavour asymmetry, suffering weak washout, contributes largely after sphaleron conversion. Thus for all benchmark points, the surviving flavours are not same reflecting the hierarchical and flavour dependent washout effects. As stated earlier, the final baryon asymmetry for each benchmark point is computed after the cancellation between the different $Y_{\Delta_\alpha}$'s, signifying the role of the flavoured nature of the leptogenesis dynamics.

\begin{figure}[htpb!]{\centering
\includegraphics[height = 10 cm, width = 12 cm]{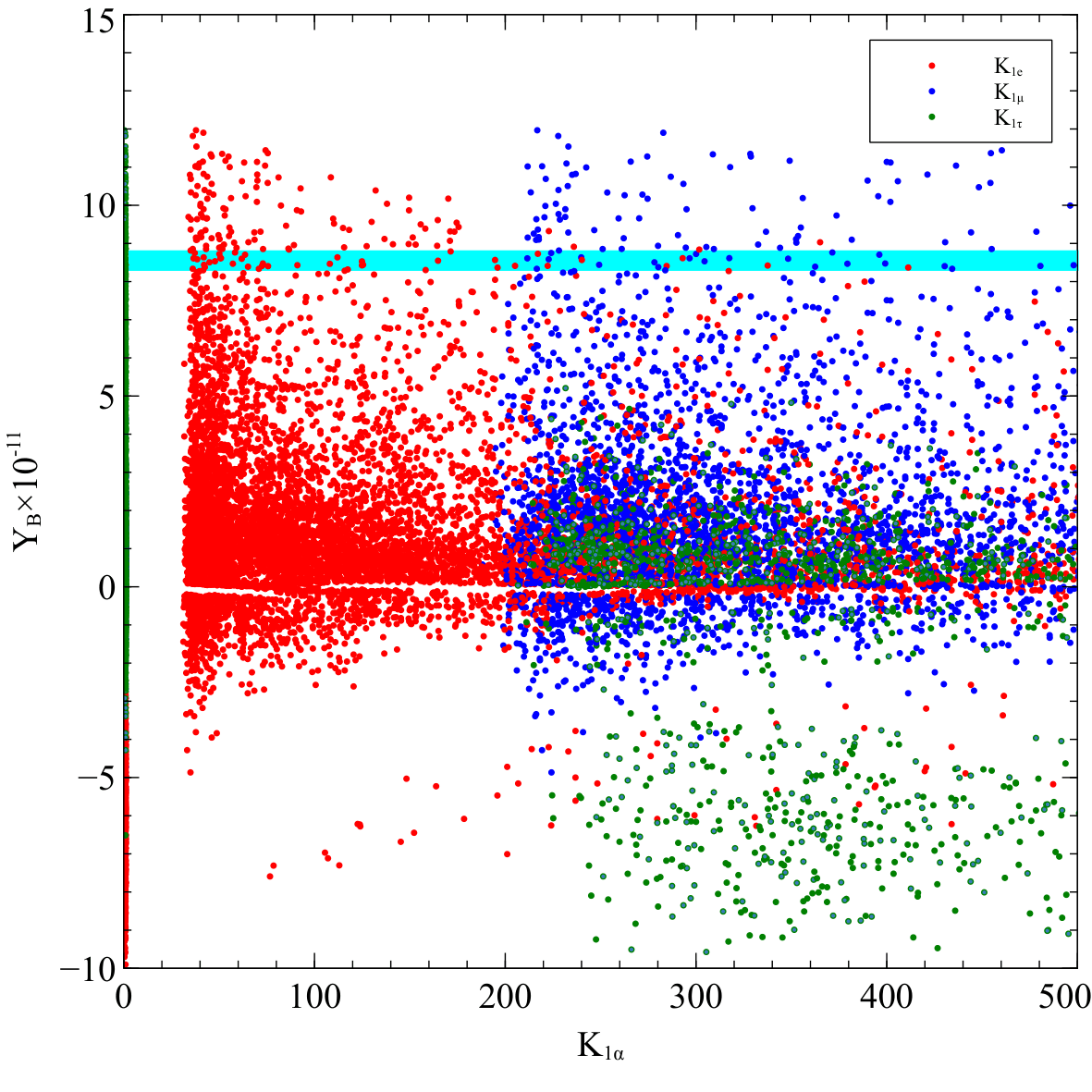}}
\caption{Variations of $Y_B$ with $K_{1\alpha}$, with $\alpha = e, \mu, \tau$. Colour coding is explained in legends.}
\label{YB-vs-washouts}
\end{figure}

To illustrate these features, we present the variation of the baryon asymmetry ($Y_B$) with respect to the individual flavour dependent washouts ($K_{1\alpha}$) in Fig.\ref{YB-vs-washouts}. It is seen that for some parameter points washout is weak ($K_{1\alpha} < 1$) in a particular flavour direction and strong for the others. The final asymmetry generated due to the three flavour directions are decided by the competition between the CP-violating source terms and the washout effects coming from inverse decay and $\Delta L = 1$ scatterings. Therefore, strong washout in one flavour direction does not necessarily ensure a smaller $Y_B$.

\subsection{Benchmark analysis}

For the representative benchmark points BP1, BP2 and BP3, which can produce the observed baryon asymmetry, we tabulate the relevant parameters in Table \ref{tab:benchmark}. We have chosen three benchmark points with low, medium and high lightest heavy neutrino masses. It has been checked that the washouts owing to $N_2$ in all the flavour channels are very strong, thereby erasing all three flavoured asymmetries for $N_2$. Thus while solving the Boltzmann equations, we only consider the evolution of $N_1$, for the points where the mass difference between $N_1$ and $N_2$ are very tiny. For all the tabulated benchmark points, the successful leptogenesis can be realised through different flavour directions, after complying with the low-energy neutrino oscillation and relevant theoretical constraints.

\begin{table}[htpb]
\centering
\begin{tabular}{|c|c|c|c|}
\hline\hline
Parameter & BP1 & BP2 & BP3  \\
\hline
$M_1$ (GeV) &16583.5 & 57566 & 97645.9  \\
$M_2$ (GeV) &16583.8 &57571.1& 97679.1  \\
$\tan\beta$ &29.52 &29.07 &  28.85 \\
$x$ (radian)& -2.865&0.282 & -2.846 \\
$y$ (radian) &  -0.031 & -0.025 &  -0.018  \\
$\xi_{\nu_R}$ &0.145   &0.139  & 0.048 \\
$\xi_{\nu_I}$ &0.028 &0.032 & 0.037  \\
$K_{1e}$ &  41.89 & 112.02 &806.66  \\
$K_{1\mu}$ & 239.82 &663.65  & 5088.11  \\
$K_{1\tau}$ & 0.30  &0.29  & 0.68  \\
$\epsilon_{1e}$ &$6.536\times10^{-8}$ &$1.092\times 10{-7}$ & $3.699\times10^{-7} $ \\
$\epsilon_{1\mu}$ & $3.281\times10^{-8}$ & $4.087\times10^{-8}$&$6.233\times10^{-8} $  \\
$\epsilon_{1\tau}$ & $3.288\times10^{-8}$&$3.970\times10^{-8}$ & $5.398\times10^{-8} $ \\
$Y_{\Delta e}$ & 4.89   $\times10^{-11} $ & 5.18   $\times10^{-11} $ & 4.76     $\times10^{-11} $  \\
$Y_{\Delta\mu}$ &   1.69  $\times10^{-12} $ & -6.31    $\times10^{-12} $ & -1.49  $\times10^{-11} $  \\
$Y_{\Delta\tau}$ &  3.57 $\times10^{-11} $ & 4.08    $\times10^{-11} $ & 5.42  $\times10^{-11} $  \\
$Y_B$ & 8.636$\times10^{-11} $ & 8.636 $\times10^{-11} $ &8.698$\times10^{-11} $  \\
$\Delta M$ (GeV) &0.3 & 5.16  & 33.25 \\
$\Gamma_{1}$ (GeV) &$1.330 \times 10^{-7}$&$ 4.410\times10^{-6}$ & $9.640\times10^{-5}$ \\
$\Delta M/\Gamma_{1}$ &$2.255\times10^6$ & $ 1.172\times10^6$&$3.449\times 10^5$  \\
$\mathrm{BR}(\mu\rightarrow e\gamma)$ &$1.66\times 10^{-33} $ &$9.98\times10^{-34} $ &$  18.24\times10^{-33}$  \\
\hline\hline
\end{tabular}
\caption{Parameter details of the three benchmark points BP1, BP2 and BP3 used in the analysis.}
\label{tab:benchmark}
\end{table}

\begin{figure}[htpb!]{\centering
\subfigure[BP1]{
\includegraphics[height = 6 cm, width = 8 cm]{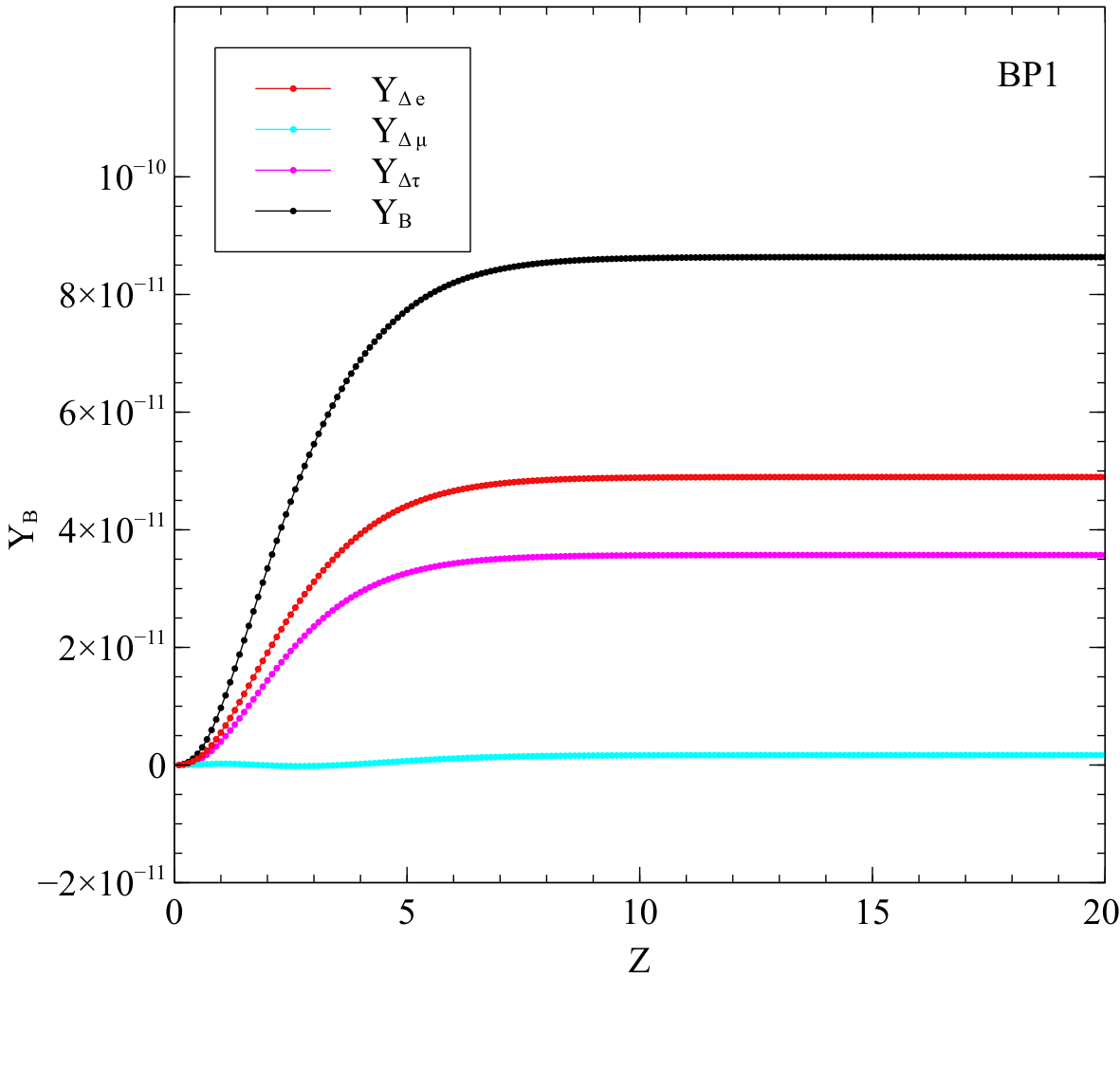}} 
\subfigure[BP2]{
\includegraphics[height = 6 cm, width = 8 cm]{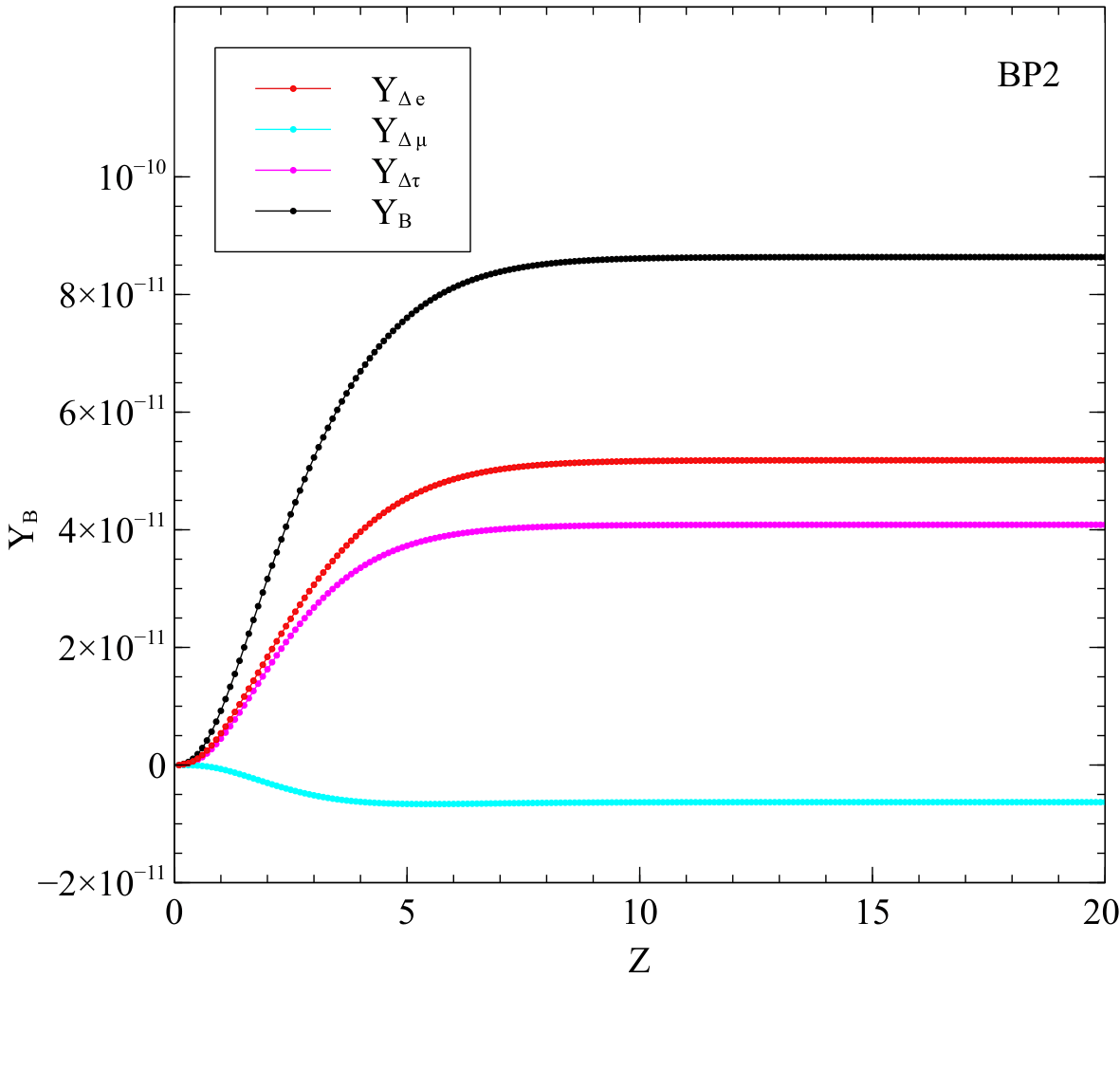}}\\
\subfigure[BP3]{
\includegraphics[height = 6 cm, width = 8 cm]{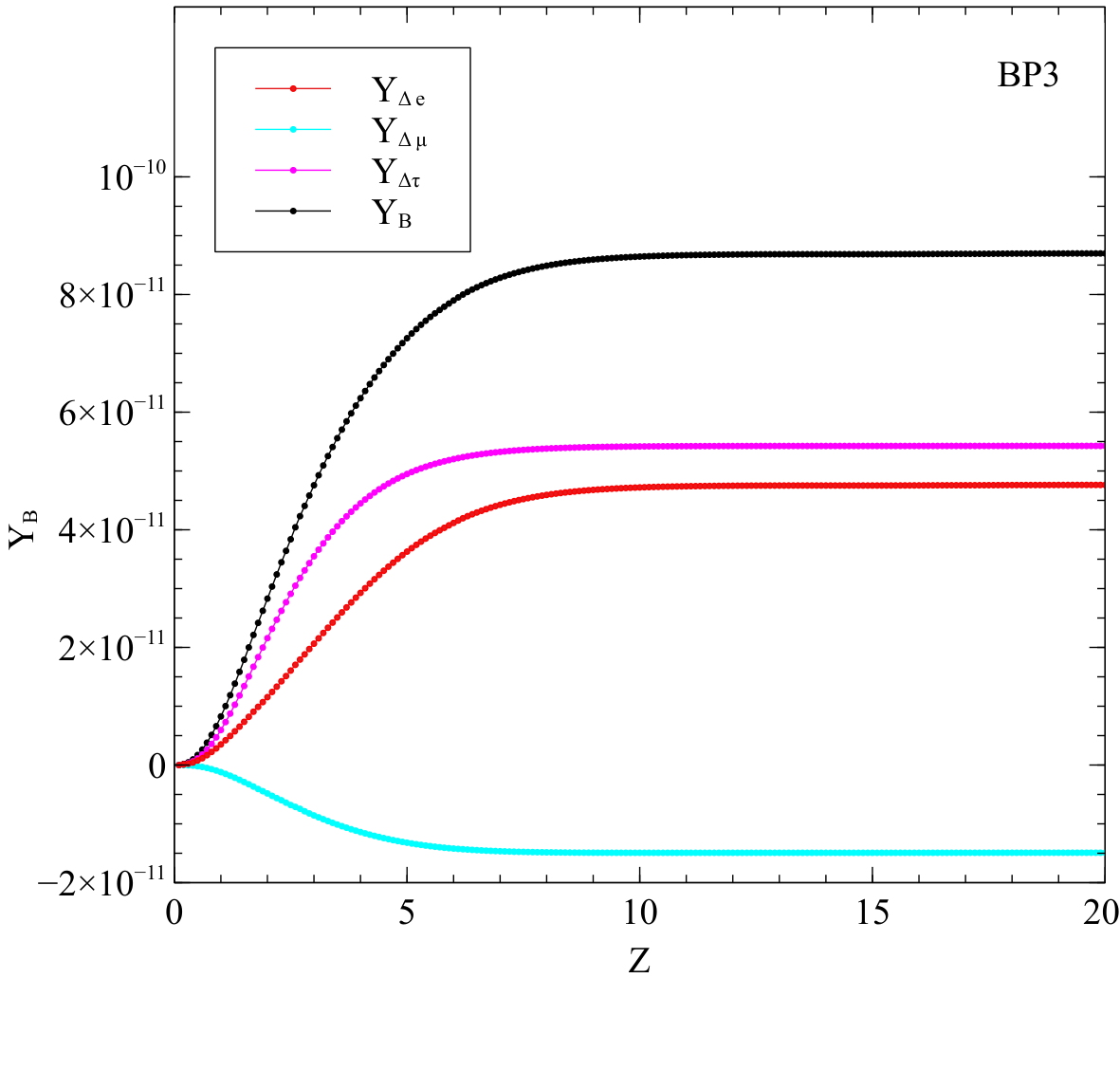}}}
\caption{Variations of $Y_{\Delta_e}$, $Y_{\Delta_\mu}$, $Y_{\Delta_\tau}$ and $Y_B$ as a function of the dimensionless variable $z=M_1/T$ for three benchmark points BP1, BP2 and BP3. Colour coding is explained in legends.}
\label{asymm-var}
\end{figure}

The evolutions of individual flavour asymmetries $Y_{\Delta_e}$, $Y_{\Delta_\mu}$, $Y_{\Delta_\tau}$ and $Y_B$ as a function of the dimensionless variable $z=M_1/T$ for selected benchmark points are depicted in Fig.\ref{asymm-var}. 
At small values of $z$, the CP-asymmetries are generated and simultaneously washed out during the inverse decays and $\Delta L=1$
scattering processes. As $z$ increases, the asymmetries reach their respective asymptotic values. In some of the flavour directions, $Y_{\Delta_\alpha}$ becomes negative after surviving through the washouts. The final baryon asymmetry is calculated considering the cancellations between $Y_{\Delta_e}$, $Y_{\Delta_\mu}$ and $Y_{\Delta_\tau}$.

\subsection{Dependence on heavy neutrino mass splitting}

\begin{figure}[htpb!]{\centering
\includegraphics[height = 10 cm, width = 12 cm]{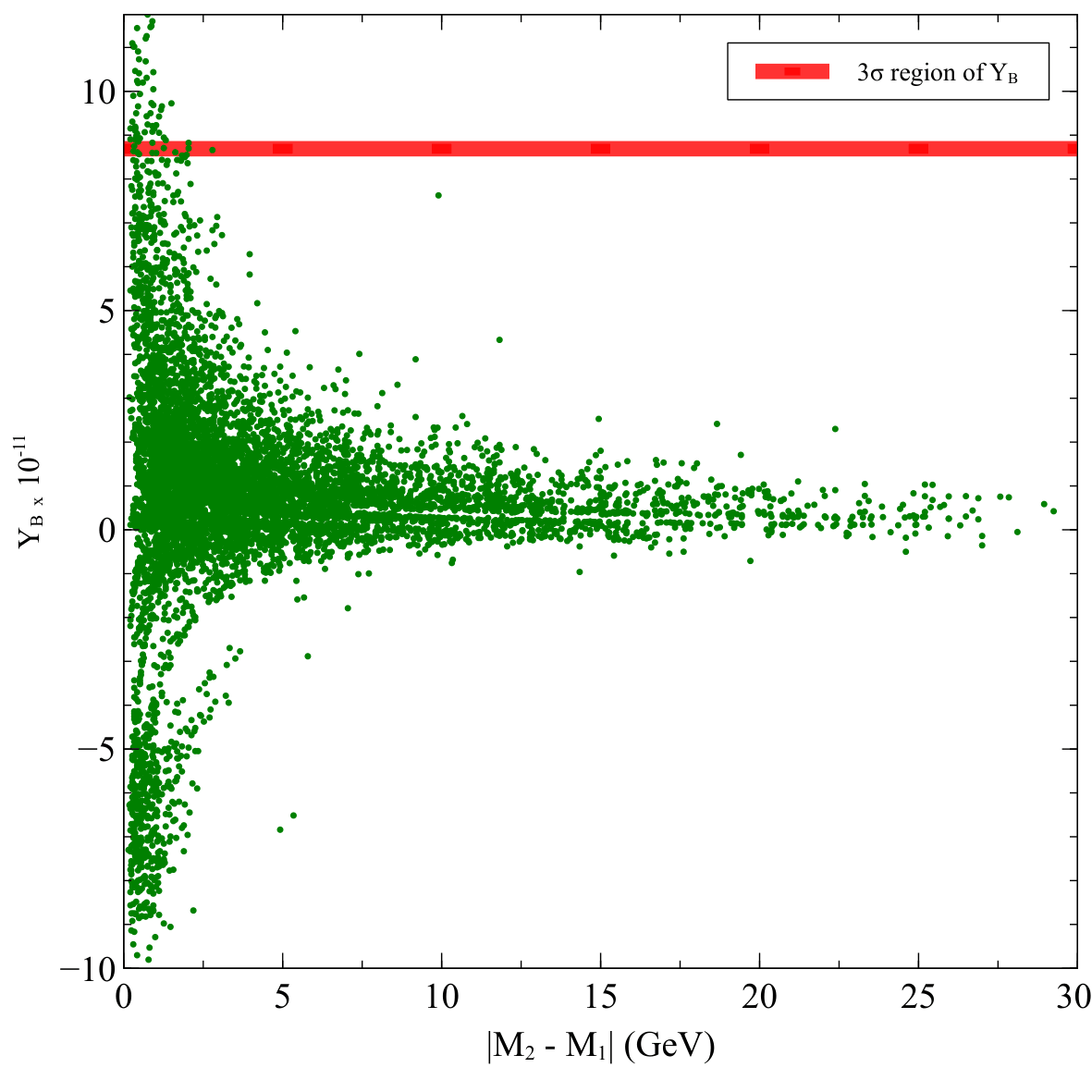}}
\caption{Variation of $Y_B$ with $\Delta M$. Colour coding is explained in legends.}
\label{YB-vs-delM}
\end{figure}

For further exploration of the dependence of the parameters on each other, we show the variation of $Y_B$ with $\Delta M = (M_2-M_1)$ in Fig.\ref{YB-vs-delM}. It is clearly evident that the points with higher magnitude of $|Y_B|$, including the points satisfying the observed baryon asymmetry are mostly crowded around the lower mass splitting region. The baryon asymmetry decreases with the increase in $\Delta M$ for most of the points. 

This type of correlation does not imply that we are exploring the resonant leptogenesis regime. For each and every parameter space point the following criterion of resonant leptogenesis is not satisfied : 
\bea
\left|M_2-M_1\right|
\lesssim
\frac{\Gamma_{N_1}+\Gamma_{N_2}}{2},
\label{resonant}
\eea
where $\Gamma_{N_1}$ and $\Gamma_{N_2}$ are the decay widths of $N_1$ and $N_2$ respectively.

Although for some points, values of $\Delta M$ are very small, the parameter points considered for the analysis, naturally belong to the non-resonant regime. For illustration, one can refer to Table \ref{tab:benchmark}. For three selected benchmark points, the condition given in Eq.(\ref{resonant}) is far from being satisfied. Therefore the distinct Yukawa structure of this model along with the role of flavoured dynamics in the generation and washouts of the asymmetries enable us to generate the observed baryon asymmetry with a wide range of TeV scale heavy neutrinos in the non-resonant regime.

\section{Conclusion}
\label{conclusion}

In this work, successful baryon asymmetry has been produced via non-resonant flavoured leptogenesis using the framework of aligned two-Higgs doublet model along with two right-handed neutrinos. In presence of the heavy right-handed neutrinos, the active neutrino mass is generated via type-I see-saw mechanism and considering the mass range of the heavy neutrinos well below $10^9$ GeV, the dynamics of the fully flavoured leptogenesis is also incorporated. The Yukawa couplings are constructed by using the Casas-Ibarra parametrisation to satisfy the neutrino oscillation data. In addition, the multi-dimensional parameter space is constrained by the perturbativity of the Yukawa couplings and the upper limit on the branching ratio of the lepton flavour violating process $\mu \to e \gamma$.

The CP-asymmetry gets contributions from the Yukawa couplings associated with both doublets. Since the three charged lepton flavours are separated in the thermal plasma, we solve the flavoured Boltzmann equations with the incorporation of decays, inverse decays, $\Delta L=1$ scatterings as well as the spectator effects. We achieve weak washout in one of the three flavour directions and strong washout in the other flavour directions. The final baryon asymmetry is generated due to the interplay of the flavoured CP-asymmetry and the hierarchical flavoured washouts. For different parameter points, the flavoured asymmetries survive in different flavour directions. 
Further we identify that the parameter points which satisfy the observed baryon asymmetry,correspond to relatively small heavy neutrino mass splittings. Although it has been checked that the mass splittings do not satisfy the conditions of resonant regime, thus pointing to the non-resonant flavoured leptogenesis. 

Therefore, the analysis explores non-resonant flavoured leptogenesis with the distinct Yukawa structure of the aligned two-Higgs doublet model along with two right-handed neutrinos, where the interplay of the generation of the flavoured lepton asymmetries and the hierarchical washouts in different flavour directions plays the crucial role in producing the observed baryon asymmetry of the universe.

\section{Acknowledgment}

IC thanks Nabarun Chakrabarty for fruitful discussions regarding numerical analysis. IC also acknowledges support
from DST, India, under grant number IFA18-PH214 (INSPIRE Faculty Award).

\appendix

\section{Lepton Flavor Violation}
\label{A}
The interaction Lagrangian involving the charged Higgs boson is given by
\beq
\mathcal{L}_{\rm int}
=
g_{s}^{ij}\,H^+\,\bar{\nu}_i\,\ell_j
+
g_{p}^{ij}\,H^+\,\bar{\nu}_i\gamma_5\ell_j
+\mathrm{h.c.}, \hspace{1cm} i,j=e,\mu,\tau\hspace{0.2cm} and \hspace{0.2cm}i\neq j
\label{Lint}
\eeq

where $g_s^{ij}$ and $g_p^{ij}$ denote the scalar and pseudoscalar couplings, respectively.

The corresponding one-loop vertex function can be written as \cite{Lindner:2016bgg}

\bea
\Gamma^\mu
&=&
-\frac{i\sigma^{\mu\nu}k_\nu}{8\pi^2m_i}
\sum_f
\Bigg[
\underbrace{
\left(
(g_{s}^{fj})^{*}g_{s}^{fi}I_{f,2}^{++}
+
(g_{p}^{fj})^{*}g_{p}^{fi}I_{f,2}^{+-}
\right)
}_{\equiv \displaystyle -(4\pi)^2A_M^{ji}}
\nonumber\\
&&
\qquad
-\gamma_5
\underbrace{
\left(
(g_{p}^{fj})^{*}g_{s}^{fi}I_{f,2}^{-+}
+
(g_{s}^{fj})^{*}g_{p}^{fi}I_{f,2}^{--}
\right)
}_{\equiv \displaystyle i(4\pi)^2A_E^{ji}}
\Bigg],~~f=1,2.
\label{Gamma}
\eea

The branching ratio for the charged lepton flavor violating decay
$\mu\rightarrow e\gamma$ is

\beq
{\rm BR}(\mu\rightarrow e\gamma)
\simeq
\frac{3(4\pi)^3\alpha_{\rm em}}
{4G_F^2}
\left(
|A_M^{e\mu}|^2
+
|A_E^{e\mu}|^2
\right),
\label{muegamma}
\eeq

where the magnetic and electric dipole amplitudes are

\bea
A_M^{e\mu}
&=&
-\frac{1}{(4\pi)^2}
\sum_f
\left[
(g_{s}^{fe})^{*}
g_{s}^{f\mu}
I_{f,2}^{++}
+
(g_{p}^{fe})^{*}
g_{p}^{f\mu}
I_{f,2}^{+-}
\right],
\nonumber\\[2mm]
A_E^{e\mu}
&=&
-\frac{i}{(4\pi)^2}
\sum_f
\left[
(g_{p}^{fe})^{*}
g_{s}^{f\mu}
I_{f,2}^{-+}
+
(g_{s}^{fe})^{*}
g_{p}^{f\mu}
I_{f,2}^{--}
\right].
\label{AEAM}
\eea

and the integrating factors
\beq
I_{f,2}^{(\pm)_1(\pm)_2}
=
\int d^3X\,
\frac{
y+(\pm)_1z\,\dfrac{m_j}{m_i}
+(\pm)_2\dfrac{m_{\nu_f}}{m_i}
}{
-xy\,m_i^2
-xz\,m_j^2
+(1-x)m_{H^+}^2
+xm_{\nu_f}^2
}.
\label{If2}
\eeq

This auxiliary function may be approximated for $m_j \ll m_i$ as

\beq
I_{f,2}^{(\pm)_1(\pm)_2}
\simeq
\frac{1}{m_{H^+}^2}
\int_0^1 dx
\int_0^1 dy\,
\frac{
x(1-x)
\left[
xy+(\pm)_2\e_f
\right]
}{
\e_f^2\lambda^2(1-x)
\left(
1-\e_f^{-2}xy
\right)
+x
}.
\label{If2approx}
\eeq

where

\beq
P_2^\pm(x)
=
x(1-x)(x\pm\e_f),
\qquad
\e_f=\frac{m_{\nu_f}}{m_\mu},
\qquad
\lambda=\frac{m_\mu}{m_{H^+}}.
\label{P2}
\eeq

\section{Notations and conventions used in Boltzmann equations}
\label{B}

The Higgs asymmetry and lepton flavour alignment coefficients $C^H $and $C^l$ respectively are given as:
\beq
C^H=\frac{8}{79}(1,\,1,\,1),
\qquad
C^\ell=\frac{1}{711}
\left(
\begin{array}{ccc}
221 & -16 & -16\\
-16 & 221 & -16\\
-16 & -16 & 221
\end{array}
\right).
\eeq

The decay reaction density is given by:

\beq
\gamma_D
=n^{\rm eq}_{N_1}
\frac{K_1(z)}{K_2(z)}
\frac{(Y^{1\dagger} Y^1)_{11}+(Y^{2\dagger} Y^2)_{11}}{8\pi}
M_1.
\eeq
where, $z=\frac{M_1}{T}.$ and $n^{\rm eq}_{N_1}=\frac{g\,M_1^3}{2\pi^2z}K_2(z)$.

The reaction density for a general $ (12\rightarrow34)$ scattering process\cite{Pilaftsis:2003gt,Pramanick:2024gvu,Giudice:2003jh} is defined like this:
\beq
\gamma_{12\rightarrow34}
=
\frac{M_N^4}{64\pi^4z}
\int_{x_{\rm min}}^\infty
dx\,\sqrt{x}\,
K_1\!\left(z\sqrt{x}\right)\hat{\sigma}(x)
\eeq
where, $x_{\rm min}
=
\frac{s_{\rm min}}{M_N^2}.$ and $s_{min}=max[(m_1+m_2)^2,(m_3+m_4)^2]$.

The reduced cross-section for s-channel and t-channel scattering processes are $\hat{\sigma}_s(x)$ and $\hat{\sigma}_t(x)$ repectively are as follows:
\beq
\hat{\sigma}_s(x)
=
\frac{3}{4\pi}
\,{\rm Tr}\!\left(Y_N^\dagger Y_N\right)
y_t^2
\left(\frac{x-1}{x}\right)^2.
\eeq

\beq
\hat{\sigma}_t(x)
=
\frac{3}{4\pi}
\,{\rm Tr}\!\left(Y_N^\dagger Y_N\right)
y_t^2
\left(\frac{x-1}{x}\right)
\left[
\frac{x-2+2a_H}{x-1+a_H}
+
\frac{1-2a_H}{x-1}
\ln\!\left(
\frac{x-1+a_H}{a_H}
\right)
\right].
\eeq

where, $y_t=\frac{\sqrt{2}\,m_t}{m_H}$ and $a_H=\frac{m_H^2}{M_N^2}.$

\bibliographystyle{JHEP}
\bibliography{ahdm} 

\end{document}